\documentclass[a4paper]{PoS}
\usepackage{bm}
\makeatletter
\def\gsim{\compoundrel>\over\sim}

\def\compoundrel#1\over#2{\mathpalette\compoundreL{{#1}\over{#2}}}
\def\compoundreL#1#2{\compoundREL#1#2}
\def\compoundREL#1#2\over#3{\mathrel
        {\vcenter{\hbox{$\m@th\buildrel{#1#2}\over{#1#3}$}}}}
\makeatother

\title{
Lambda-Nucleon and Sigma-Nucleon interactions from lattice QCD with physical masses
}

\ShortTitle{
$\Lambda N$ and $\Sigma N$ interactions from lattice QCD with physical masses
}

\author{\speaker{
Hidekatsu Nemura},$^a$
 Sinya Aoki,$^{ab}$
 Takumi Doi,$^c$
 Shinya Gongyo,$^d$
 Tetsuo Hatsuda,$^{ce}$
 Yoichi Ikeda,$^f$
 Takashi Inoue,$^g$
 Takumi Iritani,$^c$
 Noriyoshi Ishii,$^f$
 Takaya Miyamoto,$^b$
 Keiko Murano$^f$
 and 
 Kenji Sasaki$^b$\\
\llap{$^a$}
 Center for Computational Sciences, University of Tsukuba, %
 Ibaraki, 305-8577, Japan \\
\llap{$^b$}
 Center for Gravitational Physics, 
 Yukawa Institute for Theoretical Physics, Kyoto University, %
 Kyoto, 606-8502, Japan\\
\llap{$^c$}
 Theoretical Research Division, Nishina Center, RIKEN, %
 Saitama, 351-0198, Japan\\
\llap{$^d$}
 CNRS, Laboratoire de Math\'{e}matiques et Physique Th\'{e}orique, 
 Universit\'{e}de Tours, Tours, France\\
\llap{$^e$}
 iTHES Research Group, RIKEN, Saitama, 351-0198, Japan\\
\llap{$^f$}
 Research Center for Nuclear Physics (RCNP), Osaka University, 
 Osaka 567-0047, Japan\\
\llap{$^g$}
 Nihon University, College of Bioresource Sciences, 
 Kanagawa 252-0880, Japan\\

        E-mail: \email{
nemura.hidekatsu.gb@u.tsukuba.ac.jp
}
}

\abstract{
 We present our recent study on baryon-baryon ($BB$) interactions from
 lattice QCD with almost physical quark masses corresponding to
 $(m_\pi,m_K)\approx(146,525)$ MeV and large volume
 $(La)^4=(96a)^4\approx$ (8.1 fm)$^4$. In order to perform a comprehensive 
 study of $BB$ interactions based on lattice QCD calculation with almost 
 physical masses and to make better use of such 
 large scale computer resources, a large number of $BB$ interactions from
 $NN$ to $\Xi\Xi$ are calculated simultaneously. In this report, 
 we focus on the strangeness $S=-1$ channels of the hyperon
 interactions by means of HAL QCD method. The coupled-channel HAL QCD 
 method is briefly outlined. 
 The snapshots of central and tensor potentials in $^1S_0$ and $^3S_1-^3D_1$ 
 channels are presented for $\Lambda N$, $\Sigma N$ 
 (both the isospin $I=1/2, 3/2$) and their coupled-channel systems. 
 \\{\normalsize\vspace*{-53.0em}\begin{flushright}
 UTCCS-P-101,RIKEN-QHP-299,YITP-17-10
 \end{flushright}\vspace*{49.0em}}
}

\FullConference{34th annual International Symposium on Lattice Field Theory\\
		24-30 July 2016\\
		University of Southampton, UK}

\begin{document}


\section{Introduction}

Precise determination of 
the $\Lambda$-nucleon ($\Lambda N$) and 
the $\Sigma$-nucleon ($\Sigma N$) interactions 
provides a significant impact for understanding how the 
hypernuclear systems are bound. 
It has been pointed out that a $\Lambda N-\Sigma N$ coupled-channel 
interaction plays a vital role to have a hypernucleus being 
bounded\cite{Nemura:2002fu}. 
A recent experimental study shows 
a tendency to repulsive 
$\Sigma$-nucleus interaction and 
only a four-body $\Sigma$-hypernucleus ($^{4}_{\Sigma}$He) has 
been observed; 
those suggest a repulsive nature of the $\Sigma N$ interaction. 
Such understanding is useful to study properties of hyperonic matters 
inside the neutron stars, 
though a hyperonic equations of state (EOS) employed 
in such a study may contradict a recent observation of 
a massive neutron star heavier than 
$2M_{\odot}$~\cite{Demorest2010,Antoniadis:2013pzd}.

In the recent 
years, 
a new lattice QCD approach to study the hadronic interactions 
has been proposed\cite{Ishii:2006ec,Aoki:2009ji}. 
In this approach, the interhadron potential is obtained 
by means of the lattice QCD measurement of the Nambu-Bethe-Salpeter (NBS) 
wave function. 
The observables such as the phase shifts and the binding energies are 
calculated by using the resultant potential\cite{Aoki:2012tk}. 
This approach has been further extended and applied to various problems. 
See 
Refs.\cite{Sasaki:2015ifa,Ikeda:2016zwx} 
and references therein for the state-of-the-art outcomes. 
In addition, a large scale lattice QCD calculation is 
now in progress\cite{DoiIshiiSasaki2016LAT} to study the baryon 
interactions from $NN$ to $\Xi\Xi$ 
by measuring 
the NBS wave functions for 
52 channels from the $2+1$ flavor lattice QCD. 

The purpose of this report is to present 
our recent calculations of the 
$\Lambda N$ potentials 
as well as the $\Sigma N$ (both the isospin $I=1/2,3/2$) potentials 
using full QCD gauge configurations. 
Several earlier results had already been reported 
at LATTICE 2008, 
LATTICE 2009 
 and 
LATTICE 2011\cite{Nemura:2012fm} 
with heavier quark masses and smaller lattice volumes. 
This report shows the latest results of those studies, 
based on recent works reported at 
LATTICE 2013\cite{Nemura:2014eta,Nemura:2015yha};
$\Lambda N-\Lambda N$, $\Lambda N-\Sigma N$, and 
$\Sigma N-\Sigma N$ (both $I=1/2$ and $3/2$) potentials are studied 
at almost physical quark masses 
corresponding to ($m_{\pi}$,$m_{K}$)$\approx$(146,525)MeV and 
large volume $(La)^4=(96a)^4\approx$ (8.1 fm)$^4$.

\section{Outline of the HAL QCD method}

In order to study the nuclear force using the HAL QCD approach, 
we first define the equal time NBS wave function 
in particle channel $\lambda=\{B_{1},B_{2}\}$ 
with Euclidean time $t$\cite{Ishii:2006ec,Aoki:2009ji} 
\begin{equation}
 \phi_{\lambda E}(\vec{r}) {\rm e}^{-E t} = 
 \sum_{\vec{X}}
 \left\langle 0
  \left|
   B_{1,\alpha}(\vec{X}+\vec{r},t)
   B_{2,\beta}(\vec{X},t)
  \right| B=2, E, S, I 
 \right\rangle,
  \label{DefineNBSWF}
\end{equation}
where 
$B_{1,\alpha}(x)$ ($B_{2,\beta}(x)$) denotes the local interpolating field of 
baryon $B_{1}$ ($B_{2}$) 
with mass $m_{B_{1}}$ ($m_{B_{2}}$), 
and 
$E=\sqrt{k_{\lambda}^2+m_{B_{1}}^2}+\sqrt{k_{\lambda}^2+m_{B_{2}}^2}$ 
is the total energy 
in the centre of mass system of a baryon number $B=2$, strangeness $S$, 
and isospin $I$ state. 
For $B_{1,\alpha}(x)$ and $B_{2,\beta}(x)$, 
we employ the local interpolating field of octet baryons given by 
%
%
%
\begin{equation}
 \!\!\!
 \begin{array}{llll}
  p \! = \! \varepsilon_{abc} \left(
			 u_a C\gamma_5 d_b
			\right) u_c,\!
  &
  n \! = \! - \varepsilon_{abc} \left(
			   u_a C\gamma_5 d_b
			  \right) d_c,\!
  &
  \Sigma^{+} \! = \! - \varepsilon_{abc} \left(
				    u_a C\gamma_5 s_b
				   \right) u_c,\!
  &
  \Sigma^{-} \! = \! - \varepsilon_{abc} \left(
				    d_a C\gamma_5 s_b
				   \right) d_c,\!
  \\
  \Sigma^{0} \! = \! {1\over\sqrt{2}} \left( X_u \! - \! X_d \right),\!
  &
  \Lambda \! = \! {1\over \sqrt{6}} \left( X_u \! + \! X_d \! - \! 2 X_s \right),\!
  &
  \Xi^{0} \! = \! \varepsilon_{abc} \left(
                               u_a C\gamma_5 s_b
                              \right) s_{c},\!

  &
  \Xi^{-} \! = \! - \varepsilon_{abc} \left(
                                 d_a C\gamma_5 s_b
                                \right) s_{c},\!
  \\
  \mbox{where}
  &
  X_u = \varepsilon_{abc} \left( d_a C\gamma_5 s_b \right) u_c, 
  &
  X_d = \varepsilon_{abc} \left( s_a C\gamma_5 u_b \right) d_c,
  &
  X_s = \varepsilon_{abc} \left( u_a C\gamma_5 d_b \right) s_c.
 \end{array}
 \label{BaryonOperatorsOctet}
\end{equation}
%
For simplicity, we have suppressed the explicit spinor indices and 
spatial coordinates in Eq.~(\ref{BaryonOperatorsOctet}) and the 
renormalization factors in Eq.~(\ref{DefineNBSWF}). 
%
%
%
Based on a set of the NBS wave functions, we define a non-local 
potential 
%
$\left(
   \frac{\nabla^2}{2\mu_{\lambda}} + \frac{k_{\lambda}^{2}}{2\mu_{\lambda}}
  \right)
  \delta_{\lambda \lambda^{\prime}}
  \phi_{\lambda^{\prime} E}(\vec{r}) = 
  \int d^3r^\prime\, U_{\lambda\lambda^{\prime}}(\vec{r},\vec{r^{\prime}}) 
  \phi_{\lambda^{\prime} E}(\vec{r^{\prime}})$ 
with the reduced mass 
$\mu_{\lambda}=m_{B_{1}}m_{B_{2}}/(m_{B_{1}}+m_{B_{2}})$. 

In lattice QCD calculations, 
we compute the 
four-point correlation function defined by\cite{HALQCD:2012aa} 
\begin{eqnarray}
 {F}_{\alpha\beta,JM}^{\langle B_1B_2\overline{B_3B_4}\rangle}(\vec{r},t-t_0) 
 && = 
 \sum_{\vec{X}}
 \left\langle  0 
  \left|
   B_{1,\alpha}(\vec{X}+\vec{r},t)
   B_{2,\beta}(\vec{X},t)
   \overline{{\cal J}_{B_{3} B_{4}}^{(J,M)}(t_0)}
  \right|  0 
 \right\rangle,
\end{eqnarray}
where 
$\overline{{\cal J}_{B_3B_4}^{(J,M)}(t_0)}=
  \sum_{\alpha^\prime\beta^\prime}
  P_{\alpha^\prime\beta^\prime}^{(J,M)}
  \overline{B_{3,\alpha^\prime}(t_0)}
  \overline{B_{4,\beta^\prime}(t_0)}$
is a source operator that creates $B_3B_4$ 
states with the
total angular momentum $J,M$. 
The normalised four-point function 
can be expressed as
\begin{eqnarray}
 &&
  {R}_{\alpha\beta,JM}^{\langle B_1B_2\overline{B_3B_4}\rangle}(\vec{r},t-t_0) 
  =
  {\rm e}^{(m_{B_1}+m_{B_2})(t-t_0)} 
  {F}_{\alpha\beta,JM}^{\langle B_1B_2\overline{B_3B_4}\rangle}(\vec{r},t-t_0) 
  \nonumber
  \\
  \!\!\!\!&=&\!\!\!\!
   \sum_{n} A_{n}
   \sum_{\vec{X}}
   \left\langle \!\!0\!
    \left|
     B_{1,\alpha}(\vec{X}+\vec{r},0)
     B_{2,\beta}(\vec{X},0)
    \right| \!E_{n}\!\! 
   \right\rangle
   \!{\rm e}^{-(E_{n}-m_{B_1}-m_{B_2})(t-t_0)}
   \!\!+\! O({\rm e}^{-(E_{\rm th}-m_{B_{1}}-m_{B_{2}})(t-t_{0})}),
\end{eqnarray}
where $E_n$ ($|E_n\rangle$) is the eigen-energy (eigen-state)
of the six-quark system 
and 
$A_n = \sum_{\alpha^\prime\beta^\prime} P_{\alpha^\prime\beta^\prime}^{(JM)}$
$\langle E_n | \overline{B}_{4,\beta^\prime}
\overline{B}_{3,\alpha^\prime} | 0 \rangle$. 
Hereafter, the spin and angular momentum subscripts are suppressed 
for $F$ and $R$ for simplicity. 
At moderately large $t-t_0$ 
where the 
inelastic contribution 
above the pion production 
$O({\rm e}^{-(E_{\rm th}-m_{B_{1}}-m_{B_{2}})(t-t_{0})})=
O({\rm e}^{-m_{\pi}(t-t_{0})})$ 
becomes 
negligible, 
we can construct the non-local potential $U$ through 
$\left(
   \frac{\nabla^2}{2\mu_{\lambda}} + \frac{k_{\lambda}^{2}}{2\mu_{\lambda}}
  \right)
  \delta_{\lambda \lambda^{\prime}}
  F_{\lambda^{\prime}}(\vec{r}) = 
  \int d^3r^\prime\, U_{\lambda \lambda^{\prime}}(\vec{r},\vec{r^{\prime}}) 
  F_{\lambda^{\prime}}(\vec{r^{\prime}}).$ 
In lattice QCD calculations in a finite box, it is practical to use 
the velocity (derivative) expansion, 
$U_{\lambda \lambda^{\prime}}(\vec{r},\vec{r^{\prime}}) =
 V_{\lambda \lambda^{\prime}}(\vec{r},\vec{\nabla}_{r})
\delta^{3}(\vec{r} - \vec{r^{\prime}}).$ 
In the lowest few orders we have 
\begin{equation}
V(\vec{r},\vec{\nabla}_{r}) = 
V^{(0)}(r) + V^{(\sigma)}(r)\vec{\sigma}_{1} \cdot \vec{\sigma}_{2} + 
V^{(T)}(r) S_{12}
 + 
V^{(^{\ LS}_{ALS})}(r) \vec{L}\cdot (\vec{\sigma}_{1}\pm\vec{\sigma}_{2})
 + 
O(\nabla^{2}),
\end{equation}
where $r=|\vec{r}|$, $\vec{\sigma}_{i}$ are the Pauli matrices acting 
on the spin space of the $i$-th baryon, 
$S_{12}=3
(\vec{r}\cdot\vec{\sigma}_{1})
(\vec{r}\cdot\vec{\sigma}_{2})/r^{2}-
\vec{\sigma}_{1}\cdot
\vec{\sigma}_{2}$ is the tensor operator, and 
$\vec{L}=\vec{r}\times (-i \vec{\nabla})$ is the angular momentum operator. 
The first three-terms constitute the leading order (LO) potential while 
the fourth term corresponds to the next-to-leading order (NLO) potential. 
By taking the non-relativistic approximation, 
$E_{n} - m_{B_{1}} - m_{B_{2}} \simeq 
 {k_{\lambda,n}^{2} \over {2\mu_{\lambda}}} +
 O(k_{\lambda,n}^{4})$, 
and neglecting the $V_{\rm NLO}$ and the higher order terms, 
we obtain 
$\left(\frac{\nabla^2}{2\mu_{\lambda}} -\frac{\partial}{\partial t}\right)
{ R}_{\lambda\varepsilon}(\vec r,t)
\simeq 
V^{\rm (LO)}_{\lambda \lambda^{\prime}}(\vec{r}) 
\theta_{\lambda \lambda^{\prime}}
{ R}_{\lambda^{\prime}\varepsilon}(\vec r,t)$, with 
$\theta_{\lambda \lambda^{\prime}}=
{\rm e}^{( m_{B_{1}}+m_{B_{2}}-m_{B_{1}^{\prime}}-m_{B_{2}^{\prime}})(t-t_0)}$.
Note that we have introduced the matrix form 
${ R}_{\lambda^{\prime}\varepsilon} =
 \{R_{\lambda^{\prime}\varepsilon_{0}}, R_{\lambda^{\prime}\varepsilon_{1}}\}$
with linearly independent NBS wave functions 
$R_{\lambda^{\prime}\varepsilon_{0}}$ and 
$R_{\lambda^{\prime}\varepsilon_{1}}$.
%
For the spin
singlet
state, we extract the 
central potential as 
$
V_{\lambda \lambda^{\prime}}^{(Central)}(r;J=0)=
(\theta_{\lambda \lambda^{\prime}})^{-1}
({ R}^{-1})_{\varepsilon^{\prime}\lambda^{\prime}}
({\nabla^2\over 2\mu_{\lambda}}-{\partial\over \partial t})
{ R}_{\lambda\varepsilon^{\prime}}$. 
For the spin triplet state, 
the wave function 
is decomposed into 
the $S$- 
and 
$D$-wave components as 
\begin{equation}
 \left\{
 \begin{array}{l}
  R%
  (\vec{r};\ ^3S_1)={\cal P}R%
  (\vec{r};J=1)
   \equiv {1\over 24} \sum_{{\cal R}\in{ O}} {\cal R}
   R%
   (\vec{r};J=1),
   \\
  R%
  (\vec{r};\ ^3D_1)={\cal Q}R%
  (\vec{r};J=1)
   \equiv (1-{\cal P})R%
   (\vec{r};J=1).
 \end{array}
 \right.
\end{equation}
Therefore, 
the Schr\"{o}dinger equation with the LO 
potentials for the spin triplet state becomes
\begin{equation}
 \left\{\!\!\!
 \begin{array}{c}
  {\cal P} \\
  {\cal Q}
 \end{array}
 \!\!\!\right\}
 \!\!\!\times\!\!\!
 \left\{
  \!V^{(0)}_{\lambda \lambda^{\prime}}(r)
  \!+\!V^{(\sigma)}_{\lambda \lambda^{\prime}}(r)
  \!+\!V^{(T)}_{\lambda \lambda^{\prime}}(r)S_{12}
 \!\right\}
 \!\theta_{\lambda \lambda^{\prime}}
 { R}_{\lambda^{\prime}\varepsilon}(\vec{r},t-t_0)
 \!=
 \!\!\left\{\!\!\!
 \begin{array}{c}
  {\cal P} \\
  {\cal Q}
 \end{array}
 \!\!\!\right\}
 \!\!\!\times\!\!\!
 \left\{
     \!\!{\nabla^2\over 2\mu_{\lambda}} 
     -{\partial \over \partial t}
 \!\!\right\}
 \!{ R}_{\lambda\varepsilon}(\vec{r},t-t_0),
\end{equation}
from which
the 
central and tensor potentials, 
$V_{\lambda\lambda^{\prime}}^{(Central)}(r;J=0)=
(V^{(0)}(r)-3V^{(\sigma)}(r))_{\lambda\lambda^{\prime}}$ for $J=0$, 
$V_{\lambda\lambda^{\prime}}^{(Central)}(r;J=1)=
(V^{(0)}(r) +V^{(\sigma)}(r))_{\lambda\lambda^{\prime}}$, 
and $V_{\lambda\lambda^{\prime}}^{(Tensor)}(r)$ for $J=1$, can be
determined\footnote{
The potential is obtained from the NBS 
wave function at moderately large imaginary time; it would be 
$t-t_{0} \gg 1/m_{\pi} \sim 1.4$~fm. 
In addition, 
no single state saturation between the ground state 
and the excited states with respect to the relative motion, 
e.g., 
$t-t_{0} \gg (\Delta E)^{-1} = 
\left( (2\pi)^2/(2\mu (La)^2) \right)^{-1} \simeq 8.0$~fm, 
is required for the HAL QCD method\cite{HALQCD:2012aa}. 
}. 
%


\section{
Comprehensive lattice QCD calculation 
with almost physical quark masses}

$N_f=2+1$ gauge configurations at almost the physical quark masses are 
used; they are generated on $96^4$ lattice by employing the RG improved 
(Iwasaki) gauge action at $\beta=1.82$ with the nonperturbatively $O(a)$ 
improved Wilson quark (clover) action at 
$(\kappa_{ud},\kappa_{s})=(0.126117,0.124790)$ with $c_{sw}=1.11$ and 
the 6-APE stout smeared links with the smearing parameter $\rho=0.1$. 
Preliminary studies show that the physical volume is 
$(aL)^4\approx$(8.1fm)$^4$ with the lattice 
spacing $a\approx 0.085$fm and 
$(m_{\pi},m_{K})\approx(146,525)$MeV. 
See Ref.\cite{Ishikawa:2015rho} for details on the generation of 
the gauge configuration.
The periodic (Dirichlet) boundary condition is used for spacial (temporal) 
directions; wall quark source is employed with Coulomb gauge fixing which 
is separated from the Dirichlet boundary by $|t_{DBC}-t_{0}|=48$. 
Forward and backward propagation in time are combined by using the charge conjugation 
and time reversal symmetries to double the statistics. 
Each gauge configuration is used four times by using the hypercubic 
SO$(4,\mathbb{Z})$ symmetry of $96^4$ lattice. 
A large number of baryon-baryon 
potentials including the 
channels 
from $NN$ to $\Xi\Xi$ 
are studied 
by means of HAL QCD method\cite{DoiIshiiSasaki2016LAT}. 
See also Ref.\cite{Nemura:2015yha} for the thoroughgoing 
consistency check in the numerical outputs 
and comparison at various occasions 
between the UCA\cite{Doi:2012xd} and 
the present algorithm\cite{Nemura:2014eta}. 
In this report, 52 wall sources which is about a half (52/96) of 
possible statistics are used for the 207 gauge configurations 
at every 10 trajectories. 
Statistical data are averaged with the bin size 23. 
Jackknife method is used to estimate the statistical errors.

\section{Results}

\subsection{Effective masses from single baryons' correlation function}

\begin{figure}[t]
 \centering \leavevmode
 \includegraphics[width=0.46\textwidth]{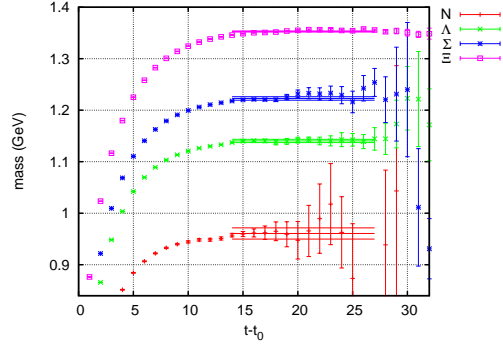}
 \caption{The effective mass of single baryon's correlation functions 
   with utilising wall sources. 
   \label{Fig_Effmass}}
\end{figure}

As mentioned above, the potential is obtained at moderately large time 
slices where the inelastic contribution 
above the pion production is suppressed. 
In addition, the single baryon's correlation functions, 
$(C_{B_{1}}(t-t_{0})C_{B_{2}}(t-t_{0}))^{-1}$, 
are used to 
obtain the normalised four-point correlation function instead of the 
simple exponential functional form ${\rm e}^{(m_{B_{1}}+m_{B_{2}})(t-t_{0})}$ 
in the actual numerical analysis.
The statistical correlation between the numerator and the denominator 
in the normalised four-point correlation function maybe beneficial to 
reduce the statistical noise. 

Fig.~\ref{Fig_Effmass} shows the effective masses of 
the single baryon's correlation function. 
The plateau starts from the time slice around $t-t_{0} \approx 14$, 
which suggests that the potentials should be obtained at 
the time slices $t-t_{0} \gsim 14$. 
However, statistics is still limited. 
In this report we present 
preliminary results 
at earlier time slices ($t-t_0=5-12$) of our on-going work. 

\subsection{Central potentials of $\Lambda N-\Sigma N$ in $^1S_0$ channel}

\begin{figure}[t]
 \begin{minipage}[t]{0.33\textwidth}
  \centering \leavevmode
  \includegraphics[width=0.99\textwidth]{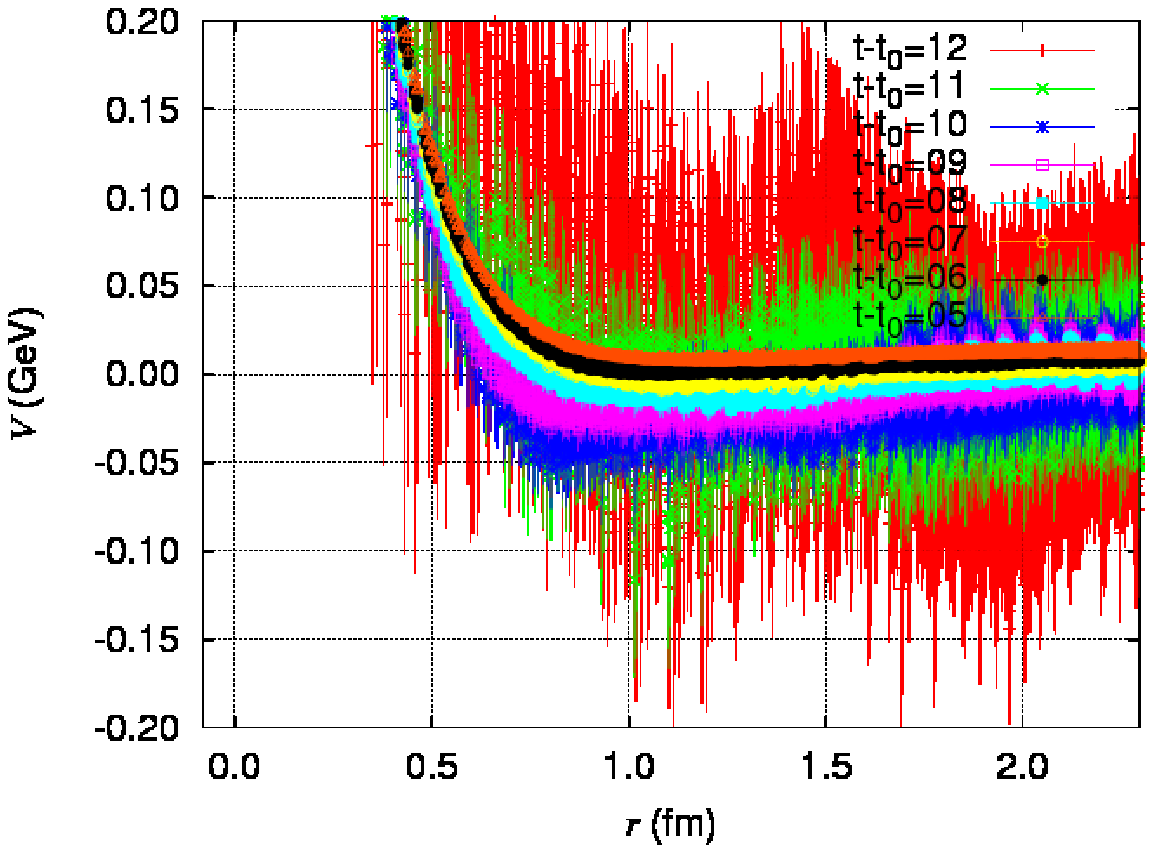}
 \end{minipage}~
 \hfill
 \begin{minipage}[t]{0.33\textwidth}
  \centering \leavevmode
  \includegraphics[width=0.99\textwidth]{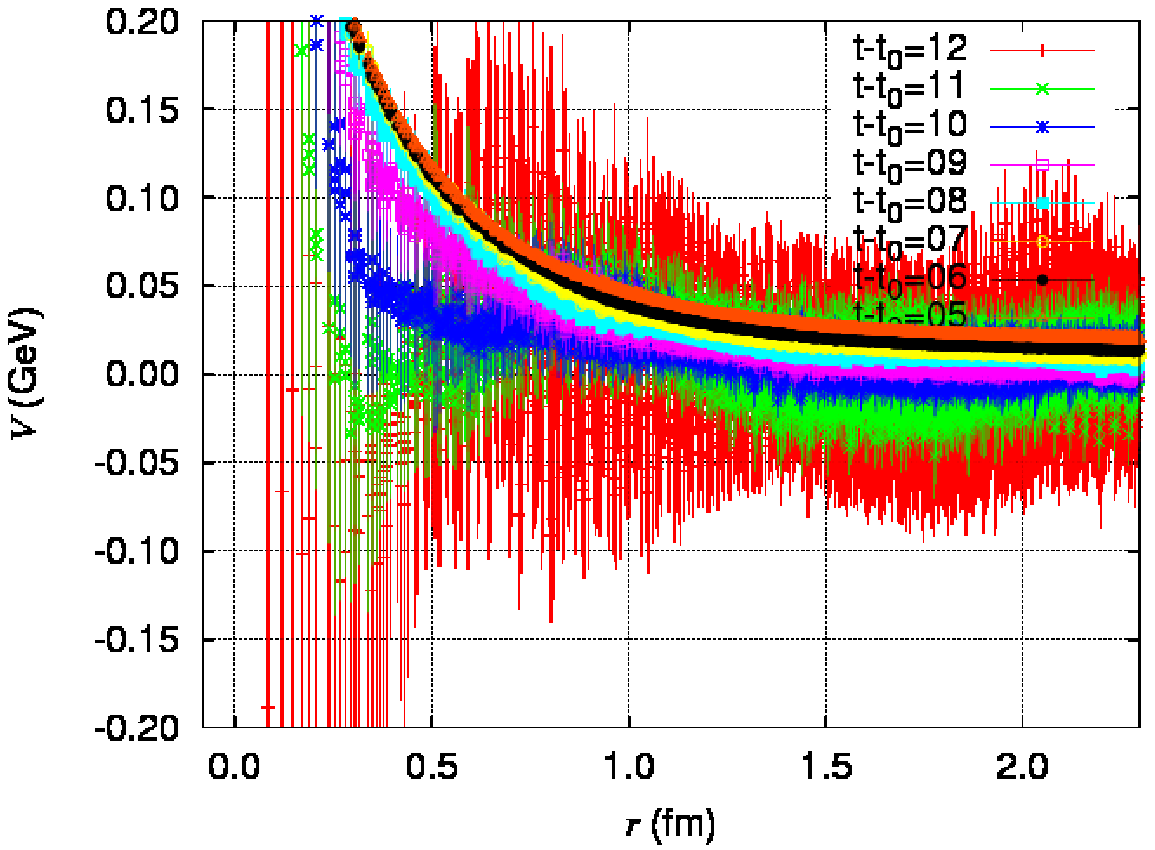}
 \end{minipage}~
 \hfill
 \begin{minipage}[t]{0.33\textwidth}
  \centering \leavevmode
  \includegraphics[width=0.99\textwidth]{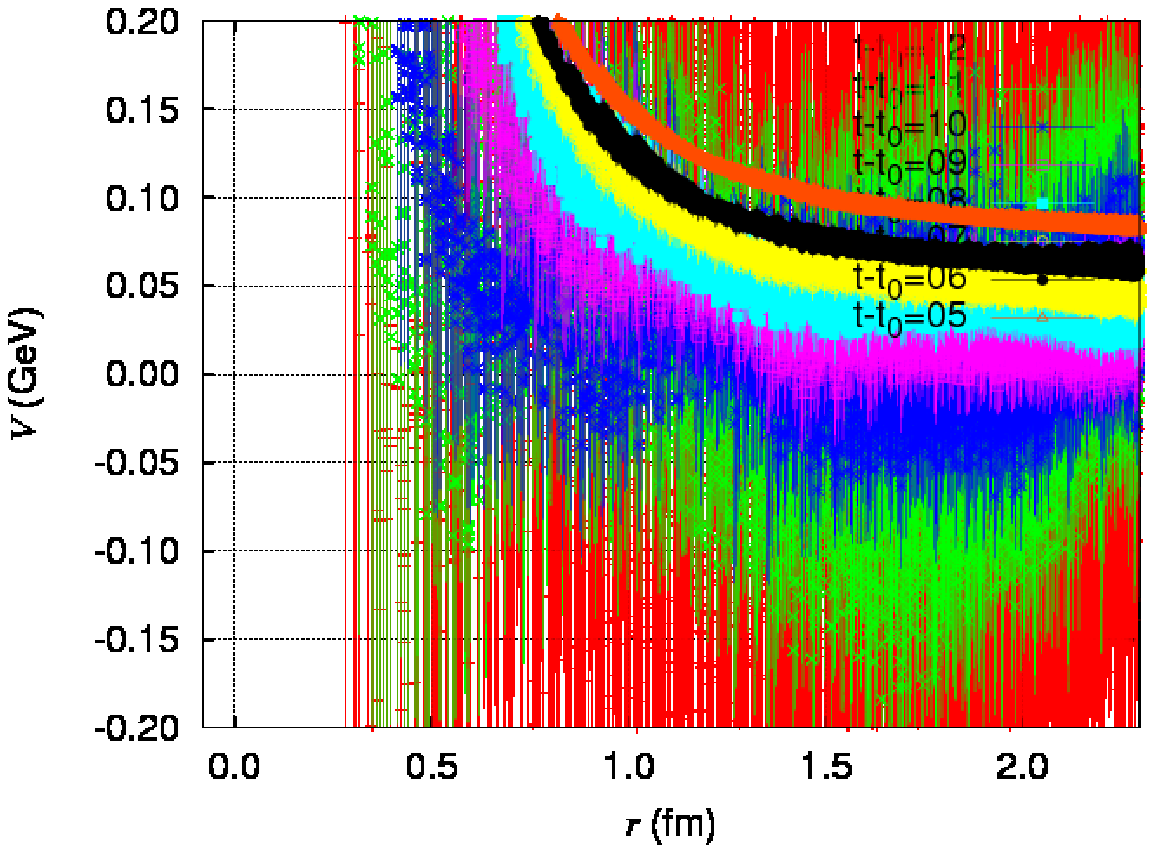}
 \end{minipage}
  \footnotesize
 \caption{Left: $\Lambda N$ central potential in the $^1S_0$ channel 
calculated with nearly physical point 
lattice QCD calculation on a volume $(96a)^4\approx$(8.1fm)$^4$ with the 
lattice spacing $a\approx 0.085$fm and 
$(m_{\pi},m_{K})\approx(146,525)$MeV. 
Centre: $\Lambda N-\Sigma N$ central potential in the $^1S_0$ channel. 
Right: $\Sigma N$ central potential in the $^1S_0$ channel. 
 \label{Fig_CentralPots1S0}}
\end{figure}

Fig.~\ref{Fig_CentralPots1S0} shows the $\Lambda N$ diagonal (left), 
$\Lambda N \rightarrow \Sigma N$ coupled-channel (centre), and 
$\Sigma N$ ($I=1/2$) diagonal (right) potentials in the $^1S_0$ channel. 
In the flavor $SU(3)$ limit, these channels are expressed in terms of 
$\bm{8}_{s}$ and $\bm{27}$ representations, 
$|\Lambda N\rangle = {1\over \sqrt{10}}( |\bm{8}_s\rangle + 3 |\bm{27}\rangle)$, and 
$|\Sigma  N\rangle = {1\over \sqrt{10}}( 3 |\bm{8}_s\rangle - |\bm{27}\rangle)$. 
Therefore the $\Lambda N$ diagonal potential is expected to be more or less
similar to the $NN$ potential in the $^1S_0$ channel. 
On the other hand, the $\Sigma N$ ($I=1/2, ^1S_0$) potential 
shows strong repulsive force 
which is consistent with the quark model's prediction. 

\subsection{Central potentials of $\Lambda N-\Sigma N$ in $^3S_1-^3D_1$ channel}

\begin{figure}[b]
 \begin{minipage}[t]{0.33\textwidth}
  \centering \leavevmode
  \includegraphics[width=0.99\textwidth]{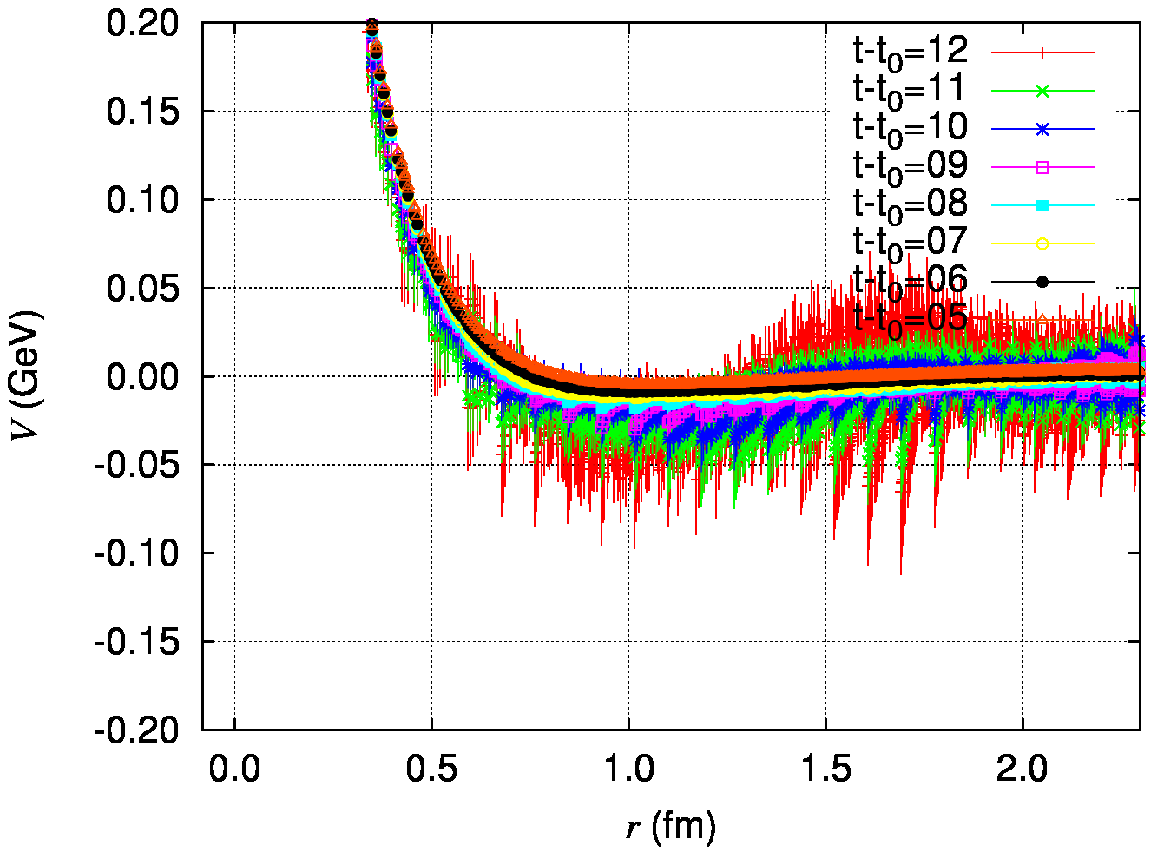}
 \end{minipage}~
 \hfill
 \begin{minipage}[t]{0.33\textwidth}
  \centering \leavevmode
  \includegraphics[width=0.99\textwidth]{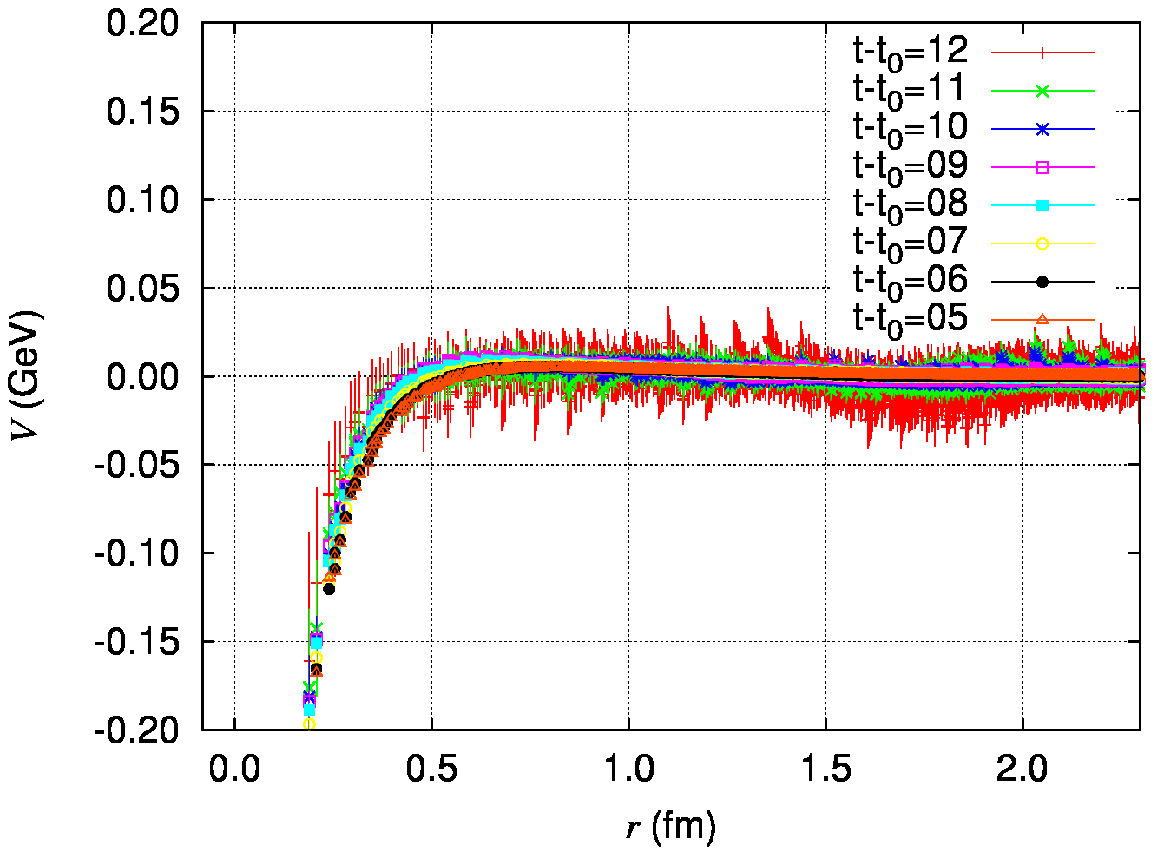}
 \end{minipage}~
 \hfill
 \begin{minipage}[t]{0.33\textwidth}
  \centering \leavevmode
  \includegraphics[width=0.99\textwidth]{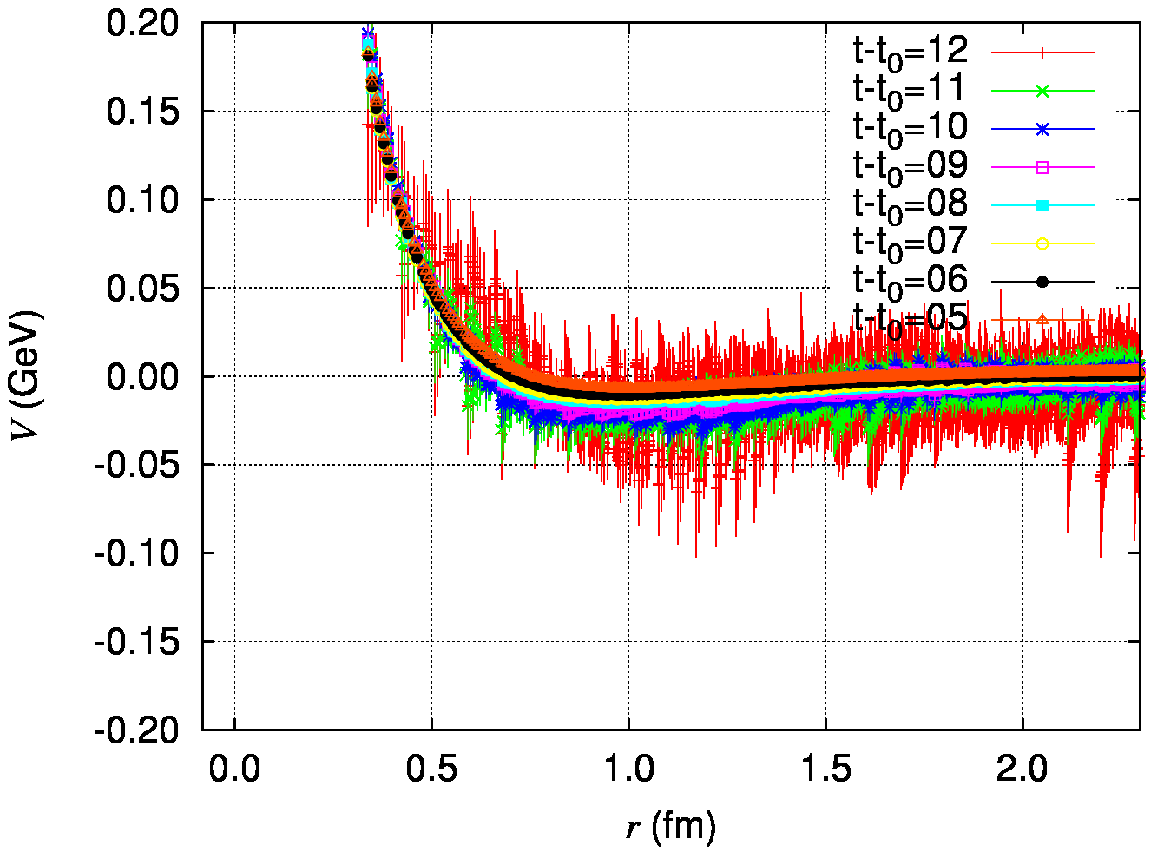}
 \end{minipage}
  \footnotesize
 \caption{Left: $\Lambda N$ central potential in the $^3S_1-^3D_1$ channel 
calculated with nearly physical point 
lattice QCD calculation on a volume $(96a)^4\approx$(8.1fm)$^4$ with the 
lattice spacing $a\approx 0.085$fm and 
$(m_{\pi},m_{K})\approx(146,525)$MeV. 
Centre: $\Lambda N-\Sigma N$ central potential in the $^3S_1$-$^3D_1$ channel. 
Right: $\Sigma N$ central potential in the $^3S_1$-$^3D_1$ channel. 
 \label{Fig_CentralPots3S13D1}}
\end{figure}

Fig.~\ref{Fig_CentralPots3S13D1} shows the $\Lambda N$ diagonal (left), 
$\Lambda N \rightarrow \Sigma N$ coupled-channel (centre), and 
$\Sigma N$ ($I=1/2$) diagonal (right) potentials in the $^3S_1-^3D_1$ 
channels. 
Relatively better signals are obtained in these states than in the 
$^1S_0$ because of the three-times larger statistics 
in the spin triplet channel. 
In the both diagonal channels, repulsive core is found and 
weakly attractive dent seems to exist in the middle distance region. 
The off-diagonal (coupled channel) potential is seen at the short distance. 

\subsection{Tensor potentials of $\Lambda N-\Sigma N$ in $^3S_1-^3D_1$ channel}

\begin{figure}[t]
 \begin{minipage}[t]{0.33\textwidth}
  \centering \leavevmode
  \includegraphics[width=0.99\textwidth]{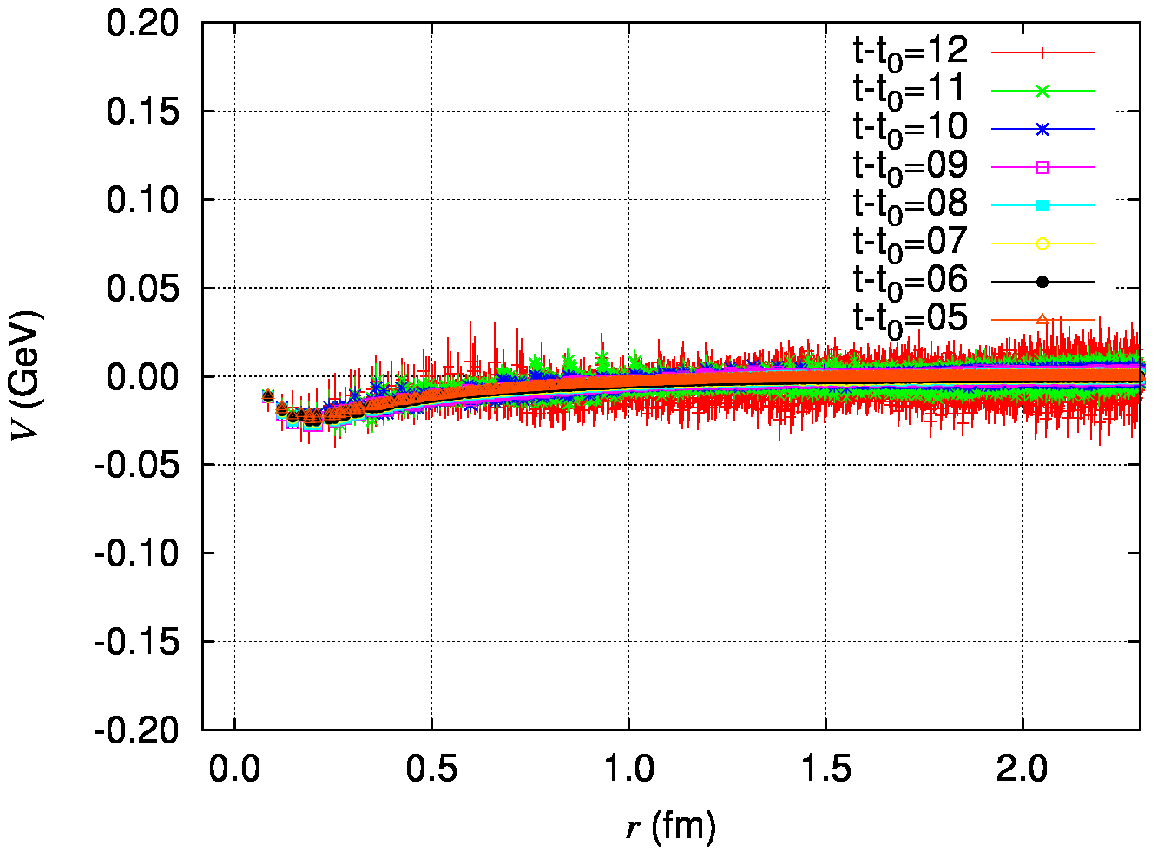}
 \end{minipage}~
 \hfill
 \begin{minipage}[t]{0.33\textwidth}
  \centering \leavevmode
  \includegraphics[width=0.99\textwidth]{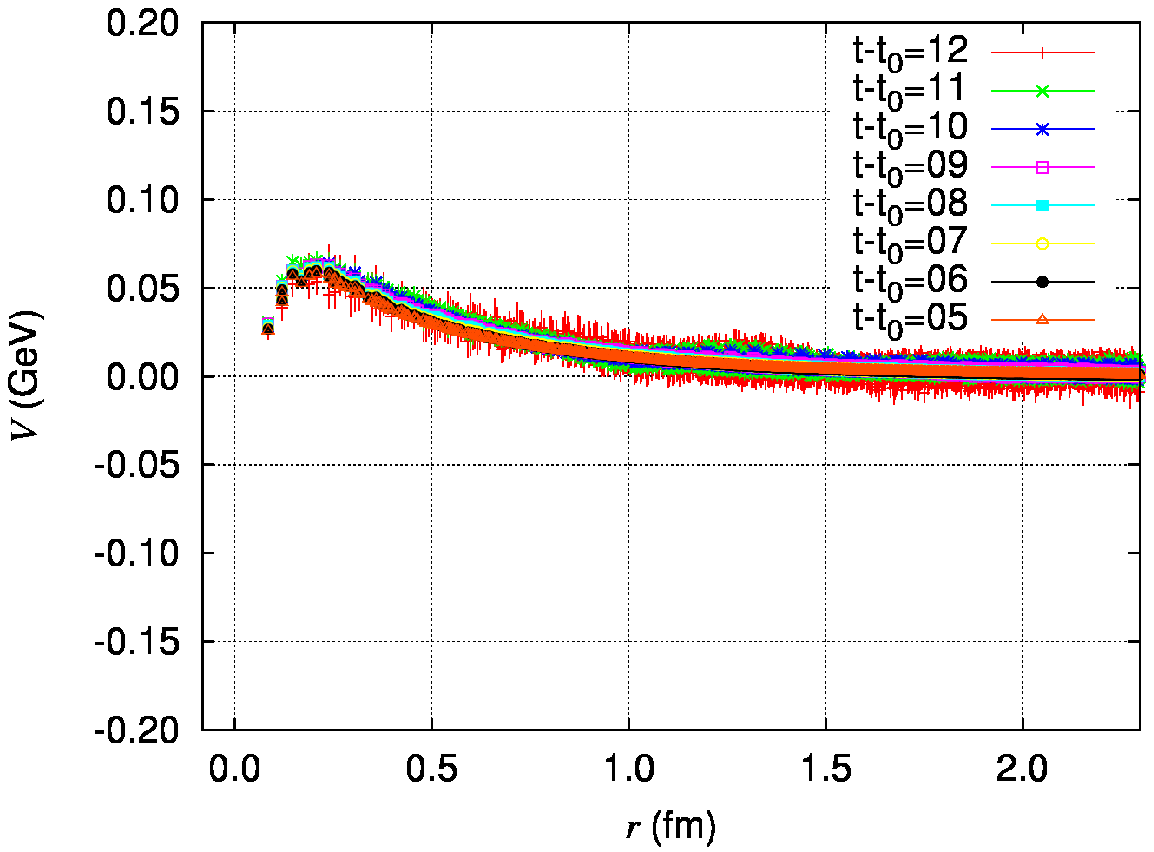}
 \end{minipage}~
 \hfill
 \begin{minipage}[t]{0.33\textwidth}
  \centering \leavevmode
  \includegraphics[width=0.99\textwidth]{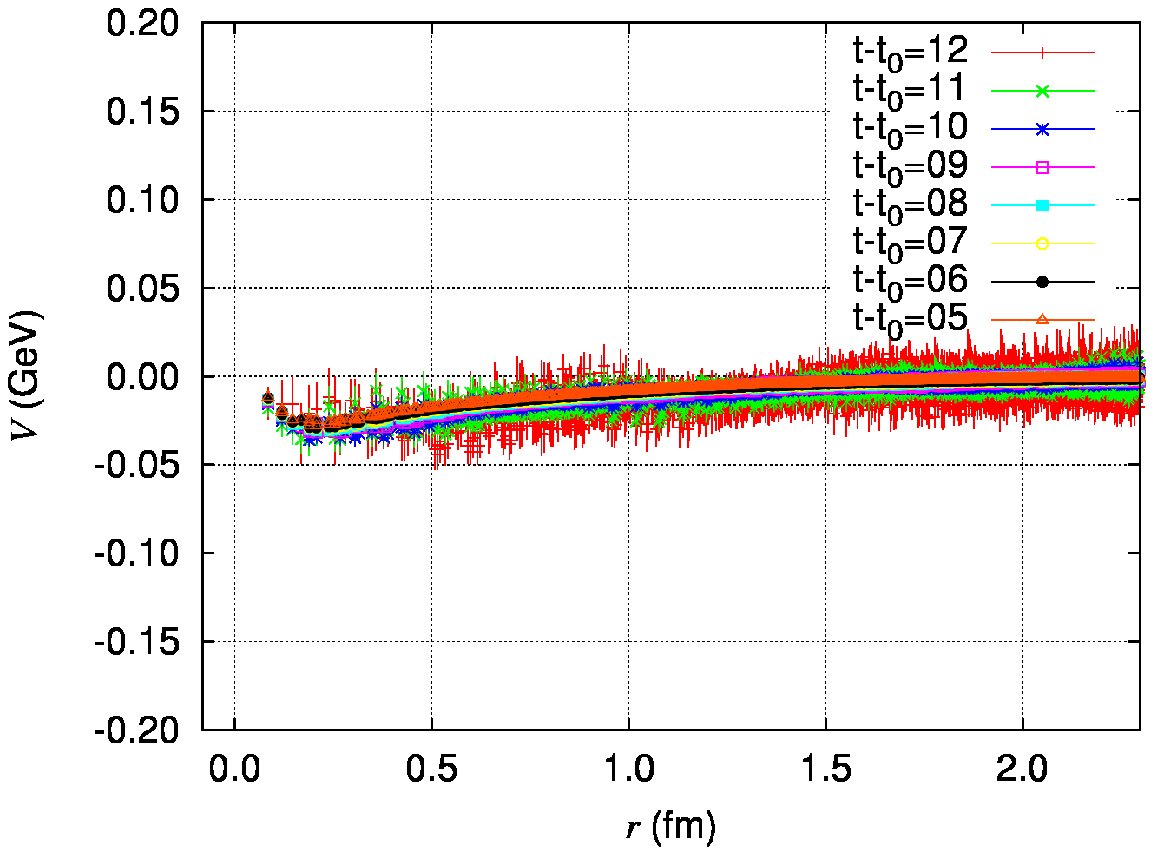}
 \end{minipage}
  \footnotesize
 \caption{Left: $\Lambda N$ tensor potential in the $^3S_1-^3D_1$ channel 
calculated with nearly physical point 
lattice QCD calculation on a volume $(96a)^4\approx$(8.1fm)$^4$ with the 
lattice spacing $a\approx 0.085$fm and 
$(m_{\pi},m_{K})\approx(146,525)$MeV. 
Centre: $\Lambda N-\Sigma N$ tensor potential. 
Right: $\Sigma N$ tensor potential. 
 \label{Fig_TensorPots3S13D1}}
\end{figure}

Fig.~\ref{Fig_TensorPots3S13D1} shows the tensor potentials in the 
$\Lambda N$ (left), 
$\Lambda N \rightarrow \Sigma N$ (centre), and 
$\Sigma N$ ($I=1/2$) (right) potentials in the 
$^3S_1-^3D_1$ channel. 
Weak tensor potentials are seen in both diagonal channels. 
Regarding the study of light hypernuclear structure~\cite{Nemura:2002fu} 
the $\Lambda N-\Sigma N$ tensor potential is expected to play 
an important role to bind one or two $\Lambda$('s) and a light nucleus. 
The present result shows that the tensor potential has 
more or less sizable strength and it is weaker than the $NN$ tensor force.

\subsection{Two central ($^1S_0$, $^3S_1-^3D_1$) and a tensor ($^3S_1-^3D_1$) potentials of $\Sigma N$ ($I=3/2$) system}

\begin{figure}[b]
 \begin{minipage}[t]{0.33\textwidth}
  \centering \leavevmode
  \includegraphics[width=0.99\textwidth]{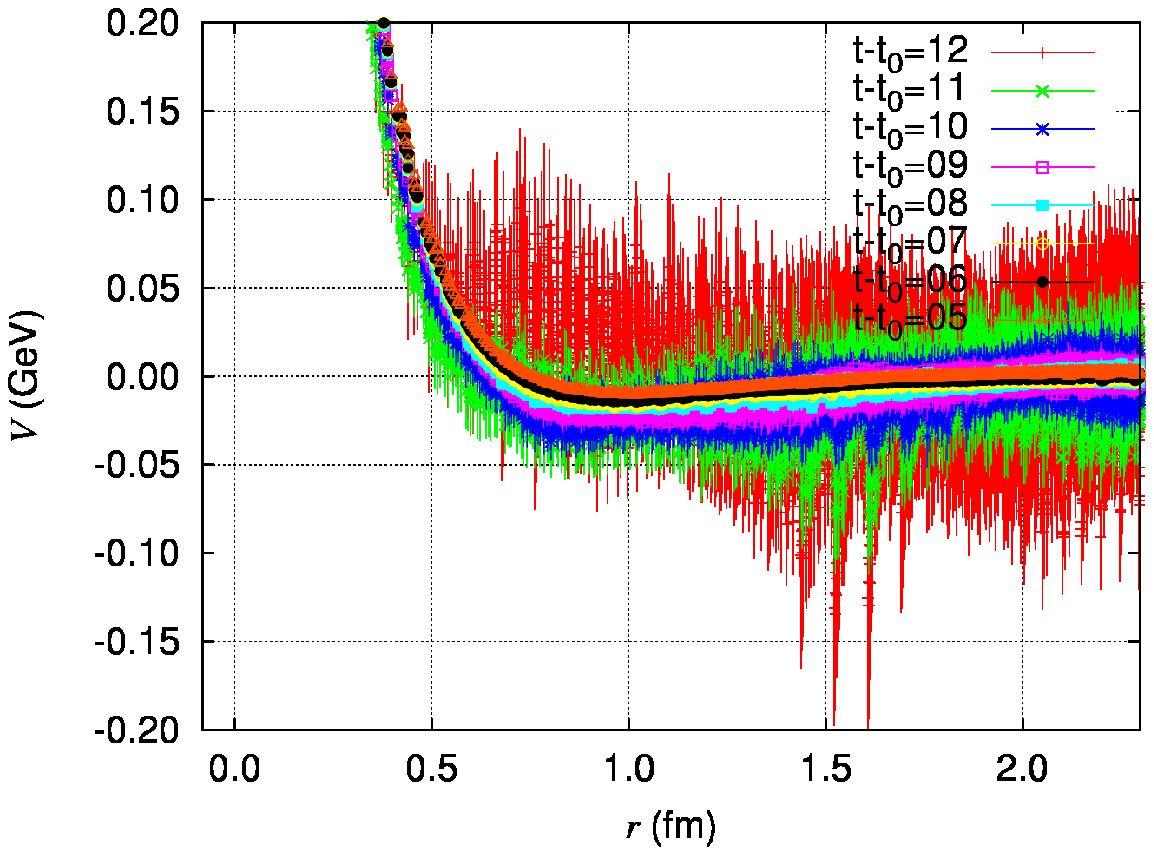}
 \end{minipage}~
 \hfill
 \begin{minipage}[t]{0.33\textwidth}
  \centering \leavevmode
  \includegraphics[width=0.99\textwidth]{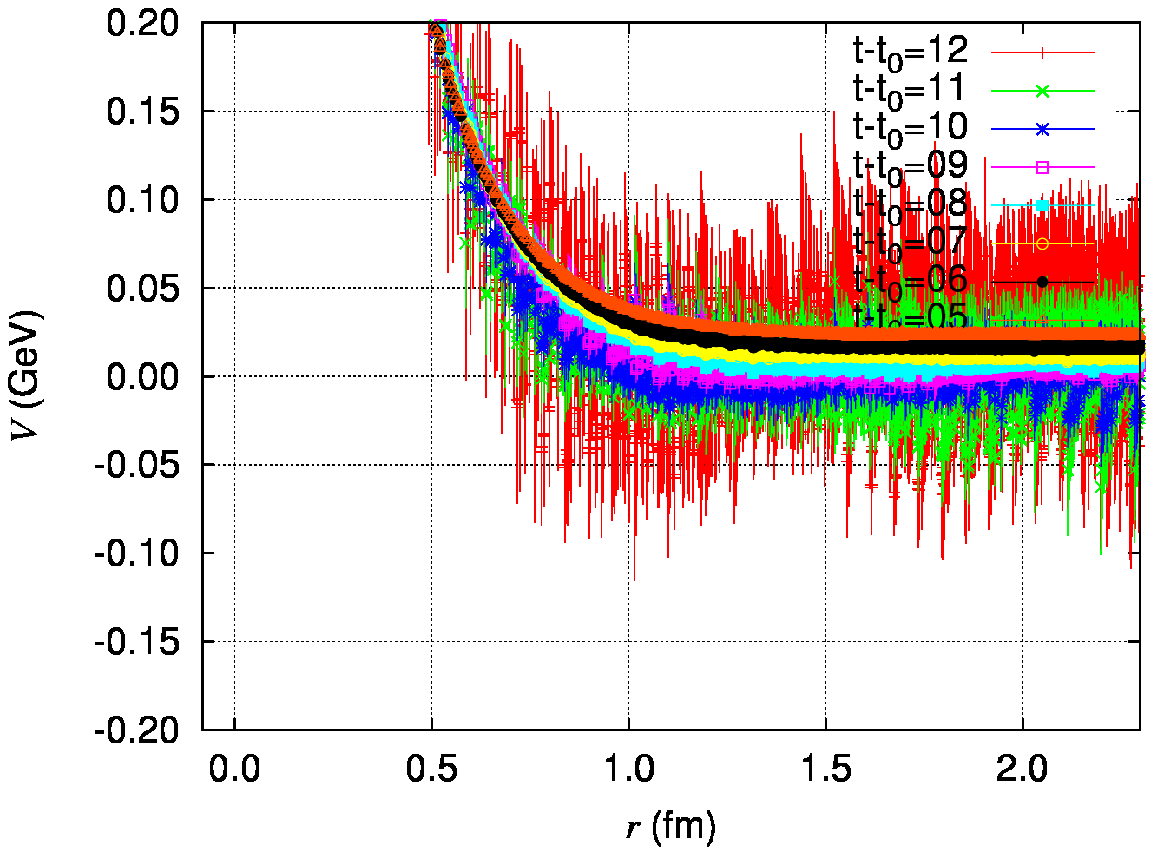}
 \end{minipage}~
 \hfill
 \begin{minipage}[t]{0.33\textwidth}
  \centering \leavevmode
  \includegraphics[width=0.99\textwidth]{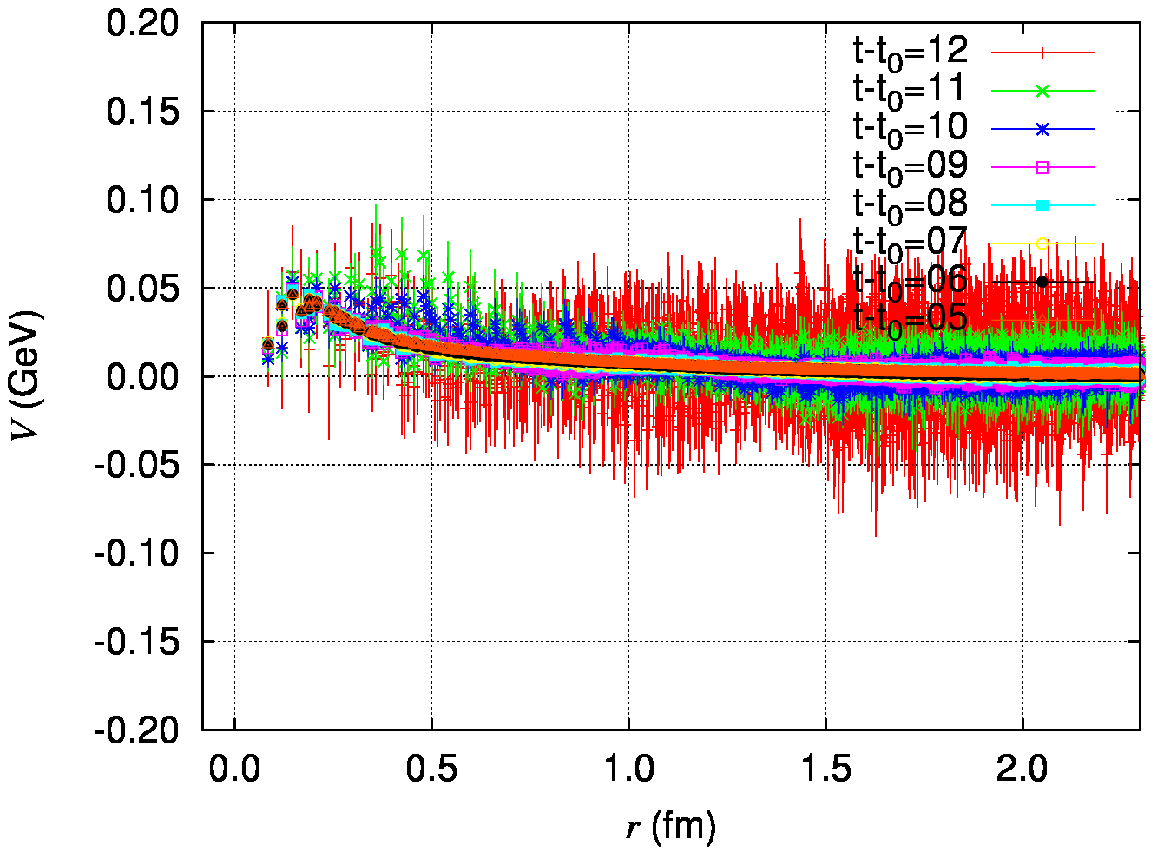}
 \end{minipage}
  \footnotesize
  \caption{The $\Sigma N$ potentials of 
    $^1S_0$ central (left), 
    $^3S_1-^3D_1$ central (centre), and 
    $^3S_1-^3D_1$ tensor  (right) 
    in the $I=3/2$ channel. 
    \label{VC1S0_VC3E1_VT3E1_SN_2I3}}
\end{figure}

Fig.~\ref{VC1S0_VC3E1_VT3E1_SN_2I3} shows the two central potentials 
in the $^1S_0$ (left) and $^3S_1-^3D_1$ (centre), and the tensor potential 
in the $^3S_1-^3D_1$ (right) channels of $\Sigma N$ ($I=3/2$) system, 
respectively. 
The $^1S_0$ $\Sigma N(I=3/2)$ channel is represented by pure $\bm{27}$ 
potential in the flavor $SU(3)$ limit 
which is same as the $^1S_0$ $NN$ potential. 
The potential shows more or less similar to the $NN$ ($^1S_0$). 
On the other hand, the $^3S_1-^3D_1$ state is represented by pure $\bm{10}$ 
irreducible representation. 
The present results seems to suggest that the central potential in the 
$^3S_1-^3D_1$ is repulsive, 
which is consistent with the quark model's prediction. 
The tensor force is also obtained. 
The lattice QCD would be a promising approach to unveil 
the origin of repulsive nature of $\Sigma N$ interaction. 

\section{Summary}

In this report, the preliminary snapshots of 
the $\Lambda N$, $\Sigma N$ and their coupled-channel potentials 
are presented. 
Both diagonal central potentials show the repulsive core 
in the short distance; 
the strengths are different from channel by channel. 
The $^1S_0$ $\Sigma N$ ($I=1/2$) and 
$^3S_1-^3D_1$ $\Sigma N$ ($I=3/2$) show relatively 
stronger repulsive cores; it is interesting to see that the 
quark model predicts similar behaviour. 
In order to obtain more clear signals for both channels, 
several efforts should be devoted to improve the following points:
(i) to increase statistics to go to larger time slices, 
(ii) to perform the analysis with taking into account the renormalization 
factors for the coupled channel potentials, 
(iii) to examine relativistic effects (i.e., higher differentials in time) 
for obtaining the potentials.

\acknowledgments

The lattice QCD calculations have been performed 
on the K computer at RIKEN, AICS 
(
hp120281, hp130023, hp140209, hp150223, hp150262, hp160211),
HOKUSAI FX100 computer at RIKEN, Wako (
G15023, G16030)
and HA-PACS at University of Tsukuba (
14a-25, 15a-33, 14a-20, 15a-30).
We thank ILDG/JLDG~\cite{ILDGJLDG}
which serves as an essential infrastructure in this study.
This work is supported in part by 
MEXT Grant-in-Aid for Scientific Research 
(16K05340, 25105505, 15K17667, 25287046, 26400281, JP15K17667),
and SPIRE (Strategic Program for Innovative Research) Field 5 project and 
``Priority issue on Post-K computer'' (Elucidation of the Fundamental Laws
and Evolution of the Universe).
We thank all collaborators in this project.

\if false
\section{...}
\fi

\end{document}